\pdfoutput=1 
\documentclass[10pt,aps,prl,amsmath,amssymb,twocolumn,letterpaper,%
         superscriptaddress,footinbib,showpacs,raggedbottom,%
         citeautoscript,floatfix]{revtex4-1}
\usepackage[bookmarks=false]{hyperref}
\usepackage{natbib}
\usepackage{amsmath}
\usepackage{amssymb}
\usepackage{bm}
\usepackage{amsthm}
\usepackage{amsfonts}
\usepackage{latexsym}
\usepackage{graphicx}
\usepackage[protrusion=true,expansion]{microtype}

\begin{document}
\title{Large isosymmetric reorientation of oxygen octahedra rotation
  axes in epitaxially strained perovskites}
\author{James M.\ Rondinelli}
  \email{jrondinelli@coe.drexel.edu}
  \affiliation{%
  X-Ray Science Division, Argonne National Laboratory, 
  Argonne, Illinois 60439, USA}
  \affiliation{Department of Materials Science \& Engineering, 
	Drexel University, Philadelphia, Pennsylvania 19104, USA}
\author{Sinisa Coh}
  \affiliation{%
  Department of Physics and Astronomy, Rutgers University, 
  Piscataway, New Jersey 08854, USA}
\date{\today}
\begin{abstract}
Using first-principles density functional theory calculations, we
discover an anomalously large bi-axial strain-induced 
octahedral rotation axis reorientation in
orthorhombic perovskites with tendency towards rhombohedral symmetry.
The transition between crystallographically equivalent (isosymmetric)
structures with different octahedral rotation magnitudes originates
from strong strain--octahedral rotation coupling available to
perovskites and the energetic hierarchy among competing octahedral
tilt patterns.
By elucidating these criteria, we suggest many functional perovskites
would exhibit the transition in thin film form, thus offering a new
landscape in which to tailor highly anisotropic electronic responses.
\end{abstract}
\maketitle
\sloppy

\begin{figure}
   \flushleft
   \vspace{-4pt}
   \includegraphics[width=0.98\columnwidth]{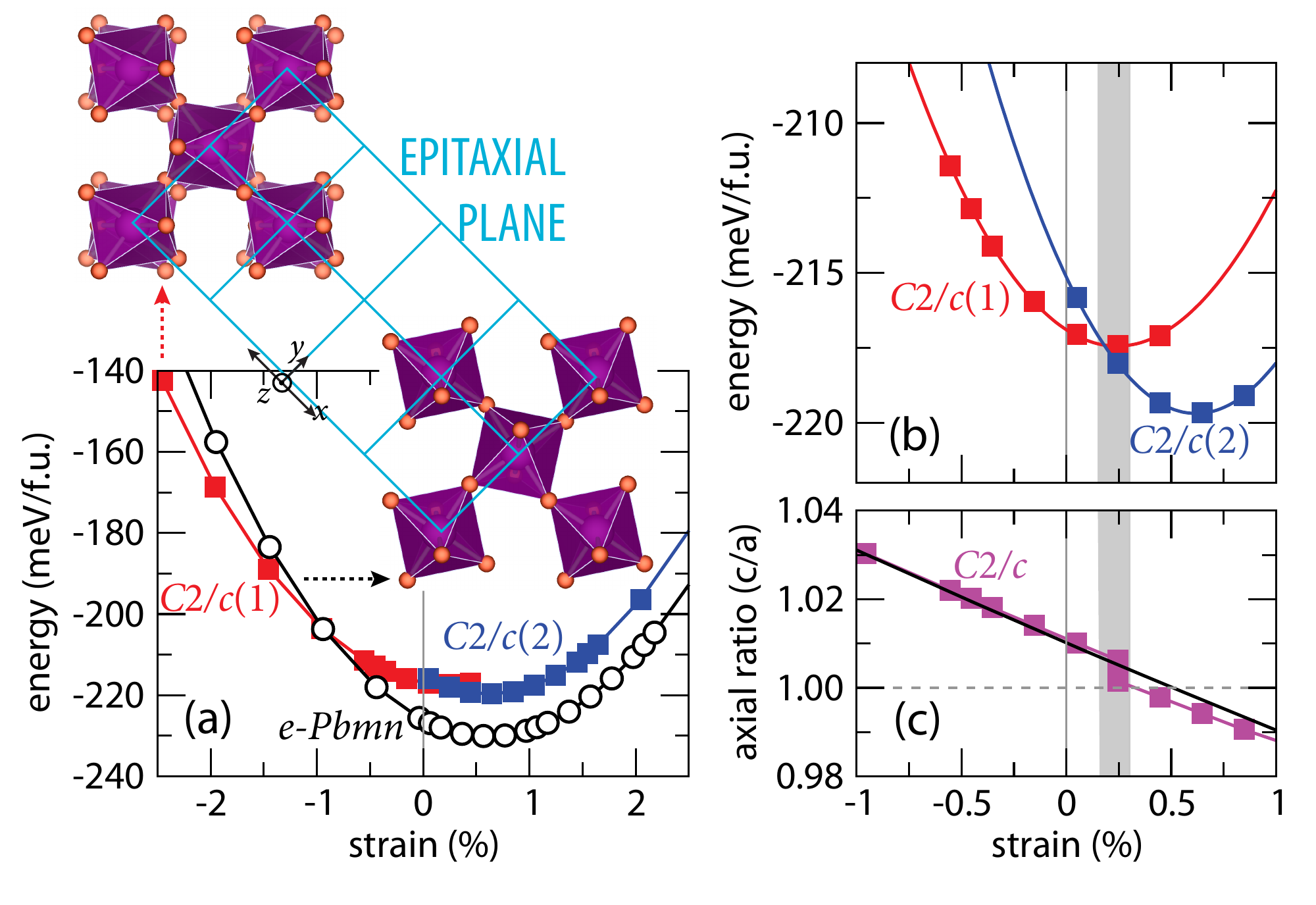}\vspace{-12pt}
   \caption{(Color) Evolution of the total energy (a) for the $e$-$Pbnm$ 
   and $C2/c$ phases with in- and out-of-phase octahedral rotations 
   (inset) along the $z$-direction.
   (b) Magnified region about the IPT (shaded). 
   %
   (c) The change in axial ratio with strain shows a
   discontinuity in the $C2/c$ phase that is absent in the 
   $e$-$Pbnm$ structure.
   \label{fig:e_strain}}
\end{figure}

\paragraph{Introduction.}
Phase transitions are ubiquitous in nature; they 
describe diverse topics ranging from crystallization and
growth to superconducting Cooper pair condensation.
Isosymmetric phase transitions (IPT)---those which show no change in
occupied Wyckoff positions or crystallographic space group---are an
intriguing class since there are relatively few examples in crystalline matter 
\cite{Cowley:1976}: 
Most condensed matter systems respond to external pressures and 
temperatures by undergoing
``conventional'' symmetry-lowering displacive \cite{Cochran:1959},
martensitic \cite{Khachaturyan:1983} or reconstructive
\cite{Toledano/Dimitrev:1996} transitions.
Furthermore, the experimental characterization and identification of a suitable 
symmetry-preserving order parameter through such transitions is often  
challenging \cite{Christy:1995}.
Although some \textit{electronic} order parameters
\cite{Lawson/Tang:1949,Caracas/Gonze:2004} that include ferroelectric
\cite{Scott:2010,Hatt/Ramesh:2009,Tinte/Rabe/Vanderbilt:2003,Tse/Klug:1988} or
orbital polarizations \cite{Friese/Grzechnik_et_al:2007} have been
proposed for IPT, which lead to subsequent changes in local cation
coordinations
\cite{Swainson/Weir:2002,Haines/Leger/Schulte:1998,Carlson/Norrestam:1998},
to the best of our knowledge, there is no case where 
the IPT connects two structures with essentially the same local bonding  
environment.
Using first-principles density functional calculations, we find an isosymmetric 
transition in the low energy rhombohedral phases of 
epitaxially strained orthorhombic perovskites and describe how to experimentally 
access it.
We show that the transition originates from non-polar distortions that describe 
the geometric connectivity and relative phase of the $B$O$_6$ octahedra 
found in perovskites.
Although a previous IPT in a thin film perovskite that relies on 
strong strain--polar phonon coupling has been 
reported \cite{Hatt/Spaldin/Ederer:2010}, 
we describe here a universal symmetry preserving transition  
that originates from the strong lattice--octahedral rotation coupling ubiquitous in 
nearly all perovskites. 
For this reason, the large isosymmetric reorientation of the oxygen rotation axes 
should be readily observable in many 
rhombohedral perovskites with diverse chemistries.
Since the dielectric anisotropy in perovskites is strongly linked to 
deviations in the octahedral rotation axis direction  \cite{Coh/Vanderbilt_etal:2010}, 
we suggest control over this transition 
could provide for highly tunable high-$\kappa$ 
dielectric actuators and temperature-free relative permittivity 
resonance frequencies \cite{Colla/Setter:1993}.
We choose LaGaO$_3$ as our model system since it has a tolerance
factor of $\tau=0.966$ indicating the perovskite structure is highly
susceptible to GaO$_6$ octahedral rotations about the principle
symmetry axes \cite{Woodward:1997b}:
At room temperature it is orthorhombic $Pbnm$ and
undergoes a first-order phase transition to rhombohedral $R\bar{3}c$
around 418~K \cite{Howard/Kennedy:1999}, with a subsequent change in
the GaO$_6$ octahedral rotation patterns from $a^-a^-c^+$ to 
$a^-a^-a^-$, respectively, in Glazer notation \cite{Glazer:1972}.
The $+$ ($-$) superscripts indicate in- (out-of)-phase rotations 
of adjacent octahedra along a given Cartesian direction.
The non-magnetic Ga$^{3+}$ cations additionally allow us to eliminate
possible contributions of spin and orbital degrees of freedom for
driving the IPT through electronic mechanisms.
\paragraph{Calculation details \& notation.}
Our density functional calculations are performed within the local 
density approximation (LDA)  as implemented in the Vienna
{\it Ab initio} Simulation Package ({\sc vasp}) 
\cite{Kresse/Furthmuller:1996a,Kresse/Joubert:1999}
with the projector augmented wave (PAW) method \cite{Blochl:1994}, 
a $5\times5\times5$ Monkhorst-Pack $k$-point mesh \cite{Monkhorst/Pack:1976} 
and a 500~eV plane wave cutoff.
We relax the atomic positions (forces to be less than 0.1
meV~\AA$^{-1}$) and the out-of-plane $c$-axis lattice constants 
for the strained films \footnote{
All strain values are given relative to the hypothetical cubic
equilibrium LDA lattice parameter (3.831~{\AA})}.
The principle difference between the ground state orthorhombic $Pbnm$ 
and metastable [12~meV per formula unit (f.u.) higher in energy] 
rhombohedral $R\bar{3}c$ phases of LaGaO$_3$  is 
that the GaO$_6$ octahedra rotate in-phase ($+$) 
along the Cartesian $z$-direction  of 
the $Pbnm$  structure while they rotate out-of-phase ($-$) 
about that same direction in the $R\bar{3}c$  structure.
Our homoepitaxial bi-axial strain calculations simulate 
film growth on a cubic (001)-terminated substrate. We therefore 
choose the $c^+$ rotations of the orthorhombic phase to be about the 
axis perpendicular to the epitaxial plane
(Fig.~\ref{fig:e_strain}),  
to evaluate the bi-axial strain effect on
the in- versus out-of-phase GaO$_6$ rotations present in the two phases.
Note, the bi-axial constraint preserves the orthorhombic 
symmetry in the $a^-a^-c^+$ phase, however, we designate the 
{\it epitaxially} ($e$) strained phase as $e$-$Pbnm$ to distinguish it from the 
bulk structure.
In contrast, the symmetry of the bulk rhombohedral
phase is lowered to monoclinic (space group $C2/c$) 
and we therefore refer to it as such
\footnote{%
In the monoclinic structures, we constrain 
the free inter-axial angle to be that of the fully relaxed
rhombohedral structure following Ref.~\onlinecite{Hatt/Spaldin/Ederer:2010}.}.

\paragraph{{Strain-stabilized structures.}}

\begin{figure*}
   \centering
   \includegraphics[width=1.86\columnwidth]{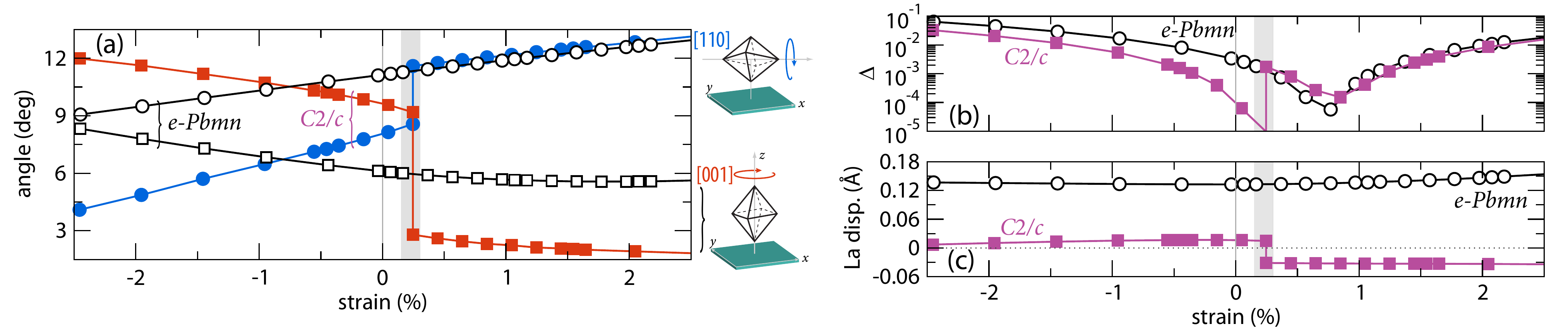}\vspace{-9pt}
   \caption{(Color) 
   Evolution in (a) the GaO$_6$ rotation angles about different directions 
   relative to the substrate, (b) 
   the octahedral distortion parameter $\Delta$, 
   and (c) the La displacements about the bulk structures  
   with epitaxial strain.
   \label{fig:xll_descriptors}}
\end{figure*}
We first compute the evolution in the total energy with bi-axial
strain for the $e$-$Pbnm$ and $C2/c$ structures
[Fig.~\ref{fig:e_strain}(a)].
We find that between approximately $-$1 to $+$3\% strain, the orthorhombic 
phase with the $a^-a^-c^+$ rotation pattern is more stable than the 
monoclinic $a^-a^-c^-$ structure.
For now we focus on the monoclinic phases [Fig.~\ref{fig:e_strain}(b)] 
near 0\% strain: 
We find an abrupt discontinuity in the first derivative of the total
energy with strain for the monoclinic structure between two states
with the same symmetry, denoted $C2/c(1)$ and $C2/c(2)$.
In contrast, we find a single continuous equation of state 
with \textit{uniform} hydrostatic pressure (over $\pm$50~GPa) for the bulk  structures.
The evolution in the $c/a$ axial ratio for these structures is 
also qualitatively different [Fig.~\ref{fig:e_strain}(c)].
The $e$-$Pbnm$ axial ratio continuously decreases with increasing
tensile strain (consistent with elastic theory), whereas in the 
$C2/c$ structures a sharp discontinuity occurs in the vicinity of $c/a\sim1$.
We find the first-order phase transition occurs at a critical strain 
of $\sim0.18$\% from intersection of the quadratic fits to the 
total energies.

\paragraph{Microscopic structure evolution.}
To investigate if the $C2/c(1) \rightarrow C2/c(2)$ transition is
indeed isosymmetric, we evaluate how the internal structural
parameters -- octhahedral tilts and bond distortions -- evolve with
epitaxial strain  [Fig.~\ref{fig:xll_descriptors}(a)].
We find a continuous evolution in the GaO$_6$ rotation angles for the 
$e$-$Pbnm$ structures (open symbols): 
the rotation axis changes from  being along the [001]-direction 
to mainly in-plane along [110] as the strain 
state changes from compressive to tensile.
In contrast, we find an abrupt change in the octahedral rotation angles  
with strain in the monoclinic phases (filled symbols).
We identify
that the $C2/c(1)$ and $C2/c(2)$ phases, despite possessing the same
symmetry are distinguishable---they have either mainly 
[110] in-plane or [001] out-of-plane 
GaO$_6$ octahedra rotations.
Consistent with the orthorhombic case 
we find that  increasing tensile strain drives the
the octahedral rotation axis into the [110]-epitaxial plane. 
The bi-axial strain is not solely accommodated by rigid octahedral
rotations.
It produces additional deviations in the Ga--O bond lengths and causes 
La cation displacements.
We quantify the former effect through the octahedral distortion parameter 
$
\Delta = \frac{1}{6}\sum_{n=1,6}[(\delta(n)-\left< \delta
    \right>)/\left< \delta \right>]^2 \, ,
$
where $\delta$ is a Ga--O bond length and $\left< \delta \right>$ is
the mean bond length in the GaO$_6$ octahedra.
With increasing strain, $\Delta$ increases, indicating that bond
stretching (and compression) occurs simultaneously 
with changes in the magnitude of the octahedral 
rotation angles to alleviate the substrate-induced 
strain [Fig.~\ref{fig:xll_descriptors}(b)].
According to our bond-valence calculations, 
the Ga--O 
bond stretching alleviates the  ``chemical strain'' imposed on the 
over-bonded Ga$^{3+}$ cations when a regular GaO$_6$ octahedra 
($\Delta \rightarrow 0$) occurs. 
The IPT allows the monoclinic phase to maintain a 
uniform charge density distribution with the $a^-a^-c^-$ tilt pattern; this is assisted by the anti-parallel La displacements [Fig.~\ref{fig:xll_descriptors}(c)], 
which change sign across the transition (shaded), maintaining a trigonal planar configuration in the GaO$_6$ rotation-created cavities. 
Note, this chemically over-bonded structure is absent in the 
$e$-$Pbnm$ structure because the $a^-a^-c^+$ tilt pattern 
($D_{2h}$ symmetry) permits non-uniform Ga--O bonds. Since the rotation 
pattern never reverses, a single La cation displacement direction occurs.

%

\begin{figure}[b]
   \centering
   \includegraphics[width=0.96\columnwidth]{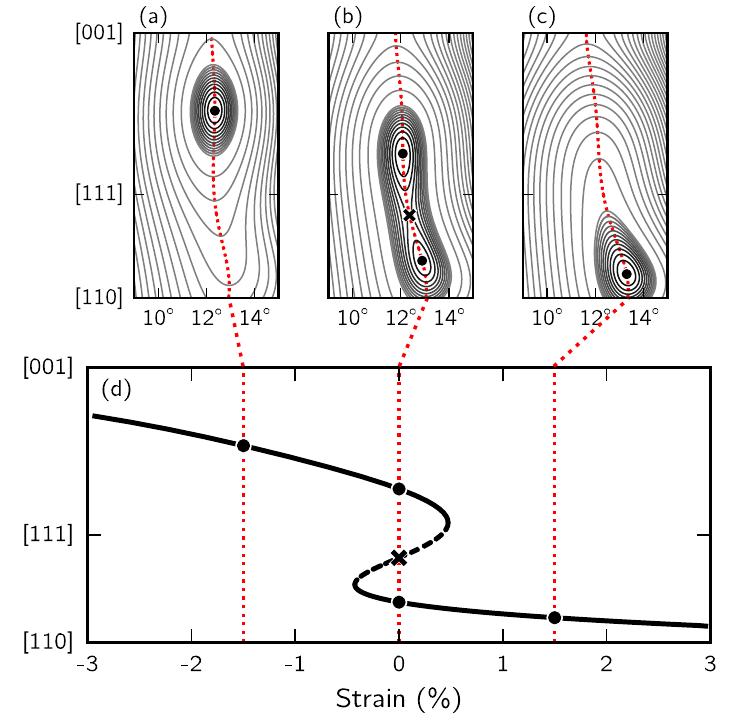}\vspace{-6pt}
   \caption{
   \label{fig:doublewell}(Color) Calculated energy (a-c) of the monoclinic 
   LaGaO$_3$ phases as a function of the GaO$_6$ rotation axis direction 
   and angle magnitude at -1.5\%, 0.0\% and 1.5\% 
   strains, respectively, with contours at 5 and 0.5~meV/f.u.\  
   (d) Position of energy minimum (solid line) or maximum (broken line) 
   with strain; circles correspond to minima (a-c), and the cross 
   indicates the saddle point in (b).}
\end{figure}

\paragraph{Origin of the IPT.}
To identify the origin of the isosymmetric transition, we first
analyze the energy of the monoclinic $a^-a^-c^-$ structures under different 
bi-axial strain states as a function of {\it direction} and
{\it magnitude} of the GaO$_6$ octahedron rotation axis.
The direction of the GaO$_6$ rotation axis with the $a^-a^-c^-$ 
pattern is constrained to be in the ($\bar{1}10$)-plane because the 
rotation pattern can be decomposed into  
$a^-a^-c^0$ and $a^0a^0c^-$ rotations with with axes aligned along 
the $[110]$- and $[001]$-directions [Fig.~\ref{fig:xll_descriptors}(a)]. 
We show in Fig.~\ref{fig:doublewell}(a-c) our first-principles results
of the energy dependence on the direction (vertical axes) and
magnitude (horizontal axes) of the GaO$_6$ rotation axis for strain
values of -$1.5\%$, $0.0\%$ and $1.5\%$, respectively.
For all strain states, we find a single well-defined energy minimum 
for each {\it direction} of the GaO$_6$ rotation axis [Fig.~\ref{fig:doublewell}(a-c)].
We therefore are able to remove the rotation angle magnitude as a
variable and to analyze the energy dependence 
solely in terms of the bi-axial strain and the GaO$_6$ rotation 
axis {\it direction} \cite{Vanderbilt/Cohen:2001}.
We show in Fig.~\ref{fig:doublewell}(d) the calculated
evolution of the extremal octahedra rotation axis directions with 
bi-axial strain:  
local energy minima (maxima) are indicated with a heavy (broken) line, and 
for -$1.5\%$, $0.0\%$ and $1.5\%$ bi-axial strains, we explicitly mark  
the extrema using symbols.
%
Consistent with our earlier structural analysis, 
we find that the rotation axis direction smoothly approaches the 
$[110]$- ($[001]$-) direction for large tensile (compressive) strains.
For the range of strains between $-0.5\%$ and $0.5\%$, we observe 
the co-existence of two energy minima separated by 
an energy maximum (broken line); this 
indicates an {\it inaccessible region} of rotation axis directions
close to $[111]$ for any value of strain and is 
consistent with a first-order transition.
Our results suggest there are two main reasons for the appearance
of the isosymmetric transition in epitaxially strained 
rhombohedral perovskites.
The first reason is that the octahedral rotations are strongly 
coupled to the bi-axial strain.
This coupling originates from the rigidity of the GaO$_6$ octahedra, since 
the rigidity causes contraction of the crystal lattice
in the direction orthogonal to the rotation axis  
\cite{May/Rondinelli:2010,Hatt/Spaldin:2010}. 
Second, the bulk rhombohedral $a^-a^-a^-$ structure of
LaGaO$_3$ is {\it higher} in energy than the bulk orthorhombic
$a^-a^-c^+$ structure. 
We now show that the energy ordering of the bulk phases 
is responsible for the {\it inaccessible region} of
rotation axis directions.
The $a^-a^-a^-$ and $a^-a^-c^+$ structures differ only in the phase of
the GaO$_6$ octahedra rotations about the $z$-axis. Thus, each 
structure can be transformed into the other through a combination of rigid
octahedral distortions. One distortion should deactivate the
$a^-$ rotation about the $z$-axis, while the other would induce the
$c^+$ rotation about the same axis.
We would expect these distortions to impose minor 
energetic penalties since they are nearly rigid 
\footnote{If a change in the $a^-a^-a^0$ rotation also occurs, 
  so as to keep the total octahedron rotation angle
  magnitude nearly constant, the energy penalty is even smaller and fully 
  consistent with the energy landscape in
  Fig.~\ref{fig:doublewell}(a-c).}.
In the present case, where the $a^-a^-a^-$ structure is higher in energy
than $a^-a^-c^+$, we expect that introduction of either of these 
distortions into the {\it higher} energy $a^-a^-a^-$ structure will 
lower the total energy %
\footnote{We calculate two unstable phonons 
	($\omega=25i$~cm$^{-1}$ and $43i$~cm$^{-1}$ at the zone-center and zone-boundary, 
	respectively) in the rhombohedral
  structure corresponding to these kind of distortions.}.
Finally, smoothness of the total energy as a function of strain and 
rotation axis direction requires 
that the difference between the number of energy 
minima ($N_{\rm min}$) and maxima ($N_{\rm max}$), for any 
value of bi-axial strain, remains fixed as elaborated in Morse
theory \cite{Vanderbilt/Cohen:2001}.
In other words, any smooth deformation which produces additional
energy maximum must also produce additional energy minimum.
Because of the first reason for the IPT mentioned above, we anticipate that
for sufficiently large compressive or tensile strains, 
the strain--octahedral rotation direction coupling dominates
to yield a single energy minimum:
$N_{\rm min}=1$, $N_{\rm max}=0$ and 
then from continuity, $N_{\rm min}-N_{\rm max}=1$ must 
remain constant for all strains. From our energetic hierarchy of the bulk structures, we conclude  that when strain induces structural distortions with 
magnitudes which nearly coincide with those of the bulk $a^-a^-a^-$ phase 
(near $0\%$ strain and $[111]$ direction), there will exist an 
energy maximum ($N_{\rm  max}=1$);  
from continuity, this must introduce two energy minima 
($N_{\rm min}=2$) at the same value of strain.
These reasons together produce the energy landscape shown 
in Fig.~\ref{fig:doublewell}(d) and require an IPT in the LaGaO$_3$ system.
For comparison, the orthorhombic phase of LaGaO$_3$ does not show an IPT 
as one varies bi-axial strain, since the second condition for the transition described 
above does not apply, ie.\ the orthorhombic structure is the global ground 
state: $N_{\rm min}$ is fixed to $1$ ($N_{\rm max} =0$)
for all strain values.
\paragraph{Accessing \& applications of the IPT.}
We obtain a $C2/c \rightarrow e$-$Pbmn$ 
transition near $-1\%$ compressive strain with respect to our 
hypothetical cubic LaGaO$_3$ phase.
In the vicinity of the IPT at 0~K, however, the $e$-$Pbmn$  
phase is the global ground state.
Although our minimal model for the IPT relies on this  
energetic ordering of the competing rotational phases 
($a^-a^-a^-$ versus $a^-a^-c^+$),   
we anticipate three experimental routes by which to access the 
essential signature of the isosymmetric transition---large 
strain-induced reorientation of the octahedral rotation axis direction.
First, the monoclinic phases could be stabilized in thin films
through the substrate coherency effect
\cite{Rondinelli/Spaldin:2010b}, where the 
film's tilt pattern adopts that of the substrate: 
Perovskite substrates with the $a^-a^-a^-$ (LaAlO$_3$)  
or the  $a^0a^0c^-$ (tetragonal-SrTiO$_3$) tilt pattern
are promising candidates.
Second, additional electronic degrees of freedom 
(first- and second-order Jahn-Teller effects), introduced through cation 
substitution, could be exploited to stabilize the IPT because they often energetically compete with the octahedra rotations \cite{Mizokawa/Khomskii/Sawatzky:1999}.
Lastly, experiments performed above the bulk LaGaO$_3$ structural 
transition temperature ($\sim100^\circ$C) would make the monoclinic phases accessible at all strains. 
%
The IPT would exhibit a weak-first order transition while still 
providing strong strain--octahedral rotation axis direction coupling.
At sufficiently high temperatures, the IPT could  be suppressed and its boundary terminated by a critical point \cite{Ishibashi/Hidaka:1991}.

We have shown that strain--octahedral rotation axis directions  are 
strongly coupled in epitaxial perovskite thin films.
%
We suggest similar large reorientations of coordinating 
polyhedra frameworks could be achieved in alternative structural families: 
thin films with the garnet, apatite or spinel structures are particularly promising.
However, the functional materials design challenge remains: how does one 
couple the rotation axis direction to additional \textit{electronic} degrees 
of freedom?
For this reason, we advocate for detailed epitaxial film studies on perovskites 
close to the  $R\bar{3}c \leftrightarrow Pnma$ phase transition 
($0.96<\tau<1.01$).
Controlling the IPT in LaCrO$_3$, LaNiO$_3$ and LaCuO$_3$ 
perovskites could yield unknown, and potentially functional, 
orbitally-, spin- and charged-ordered phases.
\paragraph{Acknowledgments.}
JMR thanks S.\ May, C.\ Fennie and L.\ Marks for discussions
and support from  U.S.\ DOE under Contract No.\ 
DE-AC02-06CH11357.
SC thanks D.\ Vanderbilt and M.\ H.\ Cohen 
for useful discussions and Rutgers-Lucent Fellowship for support.
%

%

\end{document}